\documentclass[secnumarabic, twocolumn, graphics,floatfix, superscriptaddress,tightenlines, aps, prb]{revtex4-2}
\usepackage[dvips]{graphicx}
\usepackage{amssymb,amsmath}
\usepackage{dcolumn}
\usepackage{bm}
\usepackage{color}
\usepackage[colorlinks=true, allcolors=blue]{hyperref}

\begin{document}

\title{  Resonant cooling of nuclear spins by optically-oriented holes in MAPbI$_3$ perovskite crystals 
}

\author{Mladen Kotur}
\author{Dmitri R. Yakovlev}
\author{Nataliia E. Kopteva}
\author{Bekir~Turedi}
\author{Maksym~V.~Kovalenko}
\author{Manfred Bayer}

\date{\today}




\begin{abstract}
Resonant cooling of nuclear spins by photogenerated spin-oriented holes is demonstrated for MAPbI$_3$ perovskite crystals. It is evidenced by Hanle-effect measurements under helicity-modulated excitation with variable frequency. The resonance position in magnetic field shifts toward higher fields with increasing modulation frequency. The invariance of the Hanle curve upon in-plane sample rotation is consistent with the involvement of $^{207}$Pb nuclei with spin $I=1/2$, which do not exhibit quadrupolar splitting. 
The shape of the resonance feature in the Hanle curve reveals that the nuclear spins are cooled by carriers with a negative $g$-factor, consistent with holes. The resonance fields associated with the modulation frequencies 
exceed the half-width of the weakly localized hole contribution to the Hanle curve, indicating that strongly localized holes are the primary carriers responsible for the nuclear spin cooling.


 
\end{abstract}

 \maketitle


%
The spin physics of semiconductors is contributed by the charge carrier and the nuclear spin systems as well as their interaction with each other~\cite{dyakonov2017spin}. The hyperfine interaction with optically-oriented charge carriers leads to a redistribution of the nuclear spin level populations, known as the dynamic nuclear spin polarization (DNP) effect~\cite{meier1984optical, dyakonov2017spin}. This redistribution drives the nuclear spin system out of thermal equilibrium with the lattice, thereby lowering its effective nuclear spin temperature. Such a reduction of the spin temperature is commonly referred to as nuclear spin cooling. In this regime, the nuclear spin system acquires a non-equilibrium polarization that can considerably exceed the thermal equilibrium value and persists over relatively long times due to the slow nuclear spin relaxation dynamics. To achieve the DNP, two conditions must be met: (i) the nuclear spins must be subject to a magnetic field that is non-orthogonal to the mean carrier spin, and (ii) the circularly polarized excitation light, which generates the non-equilibrium carrier spin polarization, must have a fixed helicity ($\sigma^+$ or $\sigma^-$). A standard approach for investigating the DNP relies on a magnetic field inclined relative to the incident light beam. In the presence of a tilted magnetic field $\bf B$, nuclei polarized by optically oriented carriers acquire an equilibrium average spin $\bf I$ directed along $\bf B$. The resulting nuclear polarization produces an internal magnetic field known as the Overhauser field ${\bf B}_{\rm N}$, which may either increase or decrease the effective magnetic field acting on the carrier spins. DNP has been demonstrated in classical II–VI and III–V semiconductors~\cite{lampel1968nuclear, paget1977low}, and more recently in hybrid lead halide perovskite crystals~\cite{kudlacik2024optical, kotur2026dynamic, kotur2026mapi}.

However, when the circularly polarized light is modulated between right- and left-handed helicity ($\sigma^+/\sigma^-$) at frequencies on the order of 10~kHz or higher, the rapid alternation of the carrier spin orientation strongly suppresses the nuclear spin polarization, preventing nuclear spin cooling~\citep{merkulov1980cooling}. 
While the fast helicity modulation suppresses the DNP, a resonant regime can emerge in which the periodic modulation of the optical pumping can efficiently drive nuclear spin cooling. In this regime, resonant cooling of the nuclear spins is observed under optical orientation of the carrier spins using light with alternating circular polarization that is modulated at a frequency close to the nuclear Larmor frequency in a static magnetic field applied perpendicular to the wave vector of the excitation light~\cite{meier1984optical}. Experimentally, resonant cooling of lattice nuclei can be detected by sweeping the external magnetic field perpendicular to the propagation direction of the excitation light while modulating its circular polarization. In this configuration, the resulting measurement corresponds to the Hanle curve, which describes the depolarization of carrier spin polarization in an external magnetic field~\cite{hanle1924magnetische}. If the modulation frequency matches the nuclear Larmor frequency tuned by the external magnetic field, i.e., fulfills the nuclear magnetic resonance (NMR) condition,  the Hanle curve exhibits a narrow feature  associated with the resonant cooling of the nuclear spin system. It arises from the emergence of an Overhauser field of the spin-polarized nuclei, which adds to or subtracts from the external magnetic field.

The resonant cooling can be quantitatively described within a rotating-frame model of nuclear spin cooling. It arises from the combined action of resonant pumping of the nuclear spin component precessing about the external magnetic field and its interaction with the oscillating Knight field of the optically oriented carrier spins. This leads to a net energy transfer to or from the nuclear spin system, resulting in a redistribution of nuclear spin populations toward higher or lower Zeeman energy levels, which corresponds to cooling at negative or positive spin temperatures, respectively. At resonance, a change in the phase relationship between the rotating nuclear spins and the Knight field leads to a reversal of the effective nuclear spin temperature, and consequently to a sign change of the Overhauser field. As a result, the resonant cooling produces a characteristic dispersion-shaped contribution to the Hanle curve.

Resonant cooling of the nuclear spins was experimentally observed in a wide range of II–VI and III–V semiconductor systems~\cite{kalevich1980optical, kalevich1981manifestation, zhukov2014all}. In this study, we present experimental evidence for resonant nuclear spin cooling in methylammonium lead iodide (MAPbI$_3$) perovskite crystals. From the shape of the Hanle curves and the known electron and hole $g$-factors, the resonant cooling can be attributed to the interaction of strongly localized holes with the $^{207}$Pb nuclear isotopes.

\section{Experimental results and discussion}

\begin{figure*}
\center{\includegraphics{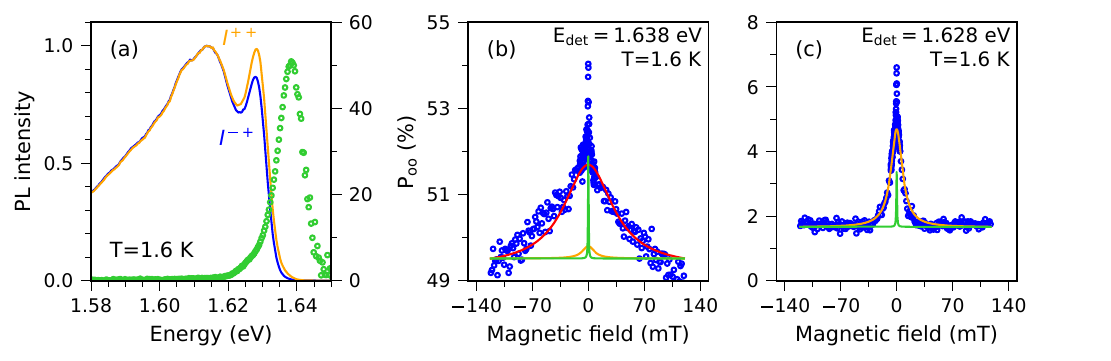}}
\caption{(a) Photoluminescence spectrum of a MAPbI$_3$ thin crystal
measured at $T = 1.6$~K. The exciting light has the photon energy of $E_{\rm exc}=1.722$~eV, the power density of $P=18$~W/cm$^2$ and is helicity-modulated between the $\sigma^+$ to $\sigma^-$ circular polarizations at $f_{\rm mod}=400$~kHz. The orange and blue lines correspond to the PL intensity under right- and left-hand circularly polarized excitation. The green open circles show the spectral dependence of the optical orientation degree. Hanle curves (circles) measured at the detection energies of (b) $E_{\rm det}=1.638$~eV and (c) $E_{\rm det}=1.628$~eV in an external Voigt magnetic field. The lines are fits according to Eq.~\eqref{eq:Hanle_formula}, showing the individual contributions from strongly localized holes (red) and electrons (orange), as well as weakly localized holes (green).}
\label{fig:spectrum}
\end{figure*}

The PL spectrum of the MAPbI$_3$ single crystal is shown in Fig.~\ref{fig:spectrum}(a). It has a maximum at 1.628~eV, which originates from the recombination of spatially separated localized carriers (electrons and holes)~\citep{wright2017band, dequilettes2019charge, kirstein2022spin, kotur2026mapi}. 
The maximum of optical orientation, about 50\%, is observed at 1.638~eV, closely corresponding to the exciton resonance $E_{\rm X}=1.636$~eV~\citep{kopteva2025optical}. It is composed of contributions from electrons and holes, as well as excitons, arising from the overlap of the recombination of localized carriers and the exciton emission at a specific energy.

When a magnetic field is applied perpendicular to the light $k$ vector, the Larmor precession of optically oriented charge carrier spins induced by this field averages out the spin component along the optical axis. The resulting decrease in circular polarization of the photoluminescence is known as the Hanle effect~\citep{meier1984optical}. Using the classical approach, the Hanle effect is expressed as
\begin{equation}
S_{z}(B_{\rm V})=\frac{S_{z}(0)}{1+(B_{\rm V}/\Delta_{\rm H})^2},
\label{eq:Hanle_formula}
\end{equation}
where $B_{\rm V}$ is the external magnetic field applied in the Voigt geometry, $\Delta_{\rm H}$ is the half-width at half-maximum (HWHM) of the Hanle curve, and $S_{z}(0)$ is the spin component along the direction of the exciting light ($z$-axis). At $B_{\rm V} = 0$, $S_{z}(0)$ equals the initial average spin of the photo-generated carriers and corresponds to the amplitude of the Hanle curve, $A_{\rm H}$. Figure~\ref{fig:spectrum} presents the Hanle curves measured at the detection energies of $E_{\rm det}=1.638$~eV shown in panel (b) and 1.628~eV shown in panel (c). The excitation beam was helicity-modulated at a frequency of 400~kHz, with the excitation energy fixed at $E_{\rm exc}=1.722$~eV. As shown in Figure~\ref{fig:spectrum}(b), for the Hanle effect detected at $E_{\rm det}=1.638$~eV, i.e., near the exciton resonance, the optical orientation degree in magnetic fields up to 125 mT does not decrease to zero and remains at approximately 50\%. This residual polarization is attributed to the optical orientation of short-lived excitons~\cite{kotur2026mapi}.

The Hanle curves exhibit three distinct contributions associated with strongly localized holes, electrons, and weakly localized holes, for details of their identification see Ref.~\onlinecite{kotur2026mapi}. The characteristic half-widths evaluated from the Hanle curve at $E_{\rm det} = 1.638$~eV are 37, 9, and 0.4~mT for the strongly localized holes, electrons, and weakly localized holes, respectively. At  $E_{\rm det} = 1.628$~eV, only two contributions are observed, from the electrons and weakly localized holes, since the amplitude of the contribution from strongly localized holes is negligible. The evaluated half-widths of 9 and 0.5~mT for the electron and weakly localized hole contributions are consistent with the values obtained at the higher detection energy.

Dynamic nuclear polarization is provided by the hyperfine-mediated transfer of angular momentum from optically oriented carriers to lattice nuclei~\citep{meier1984optical}. The DNP leads to an effective cooling of the nuclear spin system, reducing its entropy compared to thermal equilibrium with the lattice~\citep{dyakonov2017spin}. As already mentioned, when circularly polarized light is modulated between right- and left-handed helicity at frequencies of 10~kHz or higher, the nuclear spin polarization is suppressed. Nonetheless, even in the absence of a net polarization, individual nuclear spins undergo precession around the applied magnetic field with the Larmor frequency given by $\omega_{\rm L,i}=\gamma_{\rm N,i} B$. Here, $\gamma_{\rm N,i}$ is the gyromagnetic ratio of the i-th nuclear species. When the modulation frequency of the excitation light approaches the Larmor frequency of a nuclear spin in fixed external field, resonant nuclear spin cooling occurs~\citep{kalevich1980optical, kalevich1981manifestation}. This resonant cooling arises from the DNP induced by the oscillating Knight field of optically oriented electrons or holes. When the external magnetic field is scanned at a fixed modulation frequency, the resonant nuclear spin cooling manifests as a characteristic variation in the photoluminescence polarization at the field corresponding to the nuclear Larmor resonance.

\begin{figure*}
\center{\includegraphics{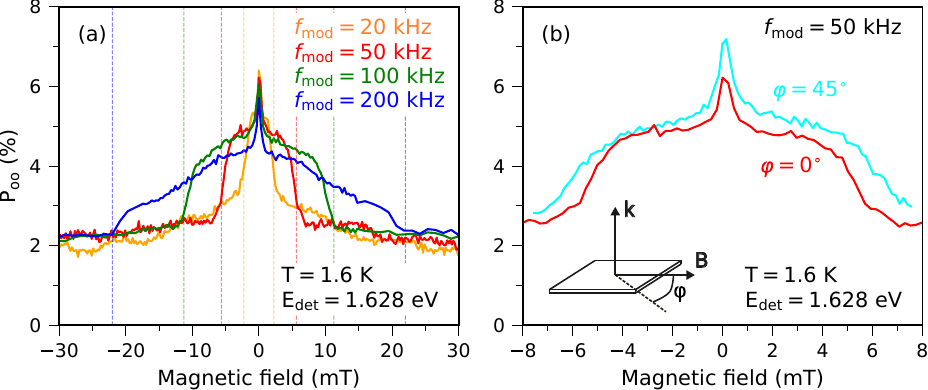}}
\caption{(a) Hanle curves measured in the Voigt geometry at $T = 1.6$~K for different modulation frequencies $f_{\rm mod}$ of the light helicity, with the in-plane angle $\varphi=0^\circ$. $E_{\rm exc}=1.722$~eV, $P=18$~W/cm$^2$, and $E_{\rm det}=1.628$~eV. The dashed vertical lines indicate the magnetic field values at which the resonance condition $f_{\rm mod}=\gamma_{\rm N} B_{\rm res}/2 \pi$ is fulfilled, where $\gamma_{\rm N}=8.882$~MHz/T is the gyromagnetic ratio of $^{207}$Pb. (b) Hanle curves measured at  $f_{\rm mod}=50$~kHz, for different in-plane angles $\varphi$.}
\label{fig:Hanle_Cooling}
\end{figure*}

Hanle curves measured at $E_{\rm det}=1.628$~eV for different modulation frequencies varying from 20 to 200~kHz are shown in Fig.~\ref{fig:Hanle_Cooling}(a), reflecting the changes induced by resonant optical cooling of the nuclear spins. It is evident that increasing the modulation frequency shifts the magnetic field at which it matches the Larmor frequency to higher values. In MAPbI$_3$, the nuclei with non-zero spin that experience hyperfine coupling with the charge carriers are $^{207}$Pb with spin $I = 1/2$ and natural abundance of 22\% and $^{127}$I with spin $I = 5/2$ and natural abundance of 100\%. The gyromagnetic ratios of $^{207}$Pb and $^{127}$I are $\gamma_{\rm N,Pb}=8.882$~MHz/T and $\gamma_{\rm N,I}=8.565$~MHz/T, respectively. These gyromagnetic ratios are close to each other, therefore, the resonant position in the Hanle curve does not unambiguously indicate which isotope is being cooled. 

Nuclei with spin $I=1/2$, such as $^{207}$Pb, do not possess a nuclear quadrupole moment and, therefore, are not subject to quadrupolar splitting. In contrast, nuclei with spin $I>1/2$, such as $^{127}$I, possess a finite quadrupole moment and experience quadrupolar splitting in the presence of electric field gradients. The resulting quadrupole-split doublets are characterized by strongly anisotropic effective $g$-factors, leading to a pronounced dependence of the resonance frequency on the sample orientation. 
%
Figure~\ref{fig:Hanle_Cooling}(b) compares the Hanle curves measured at $E_{\rm det}=1.628$~eV with the excitation light modulated at $f_{\rm mod}=50$~kHz, alternating between $\sigma^+$ and $\sigma^-$ polarization, for the sample rotated in-plane by $\varphi = 0^\circ$ and $45^\circ$. The nearly identical Hanle curves for $\varphi = 0^\circ$ and $45^\circ$ demonstrate that the resonance condition is insensitive to the in-plane sample orientation, consistent with the cooling of $^{207}$Pb nuclei. In contrast, quadrupole-split $^{127}$I nuclei are expected to exhibit a pronounced angular dependence of the resonance frequency.

\begin{figure*}
\center{\includegraphics{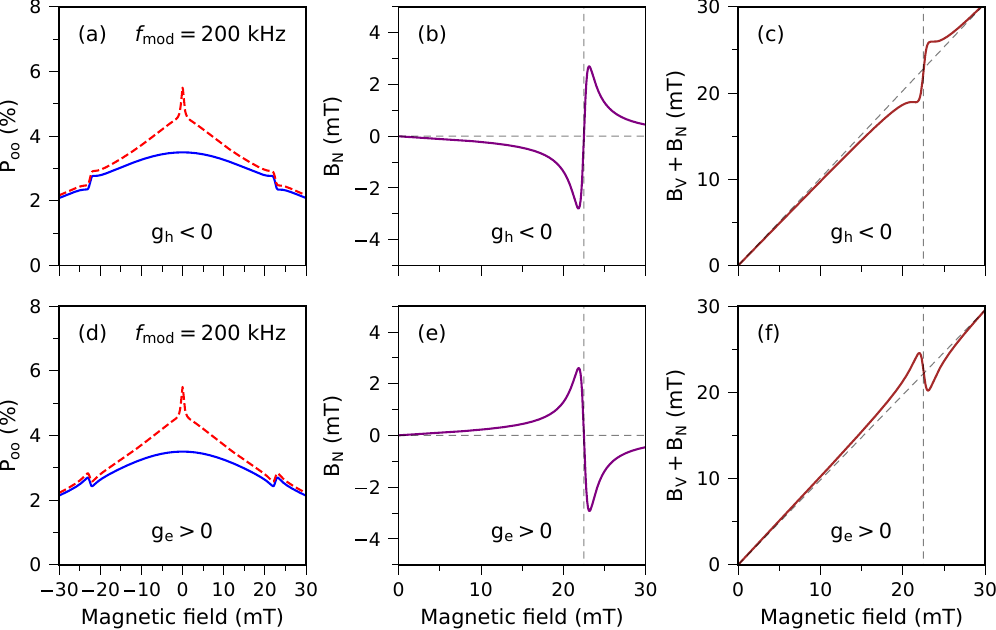}}
\caption{Modeling of the Hanle curves in the Voigt geometry under resonance cooling conditions for the helicity modulation frequency of 200~kHz. (a,d) Hanle curves calculated using Eqs.~\eqref{eq:Hanle_formula} and~\eqref{eq:nuclearfield}  using the parameters $b_{\rm N}T_1/T_{\rm 1e(h)}=200$ mT, $b_{\rm e(h)}=20$~mT, and $\Delta_{\rm H}=37$~mT for holes with $g_{\rm h}<0$ and electrons with $g_{\rm e}>0$ (blue line). The red dashed line corresponds to the same Hanle model with additional contributions from electrons and weakly localized holes with 9 and 0.4~mT half-widths, respectively. (b,e) Magnetic field dependence of the Overhauser field $B_{\rm N}$ corresponding to the Hanle curves (blue line) in panels (a) and (d). (c,f) The resulting total magnetic field, equal to the sum of the external magnetic field $B_{\rm V}$ and the Overhauser field $B_{\rm N}$, ($B_{\rm V}+B_{\rm N}$). The vertical dashed line indicates the resonance field.}
\label{fig:Hanle_Cooling_Theory}
\end{figure*}
The steady-state nuclear polarization is governed by the competition between the dynamic nuclear polarization and the spin-lattice relaxation, arising from hyperfine interaction with charge carriers as well as other relaxation mechanisms. Accounting for both the polarization and relaxation processes, the nuclear field $B_{\rm N}$ along the $x$ axis is described by the following balance equation~\cite{kavokin2026resonant}
\begin{widetext}
\begin{equation}
B_{\rm N}
=
\frac{T_1}{T_{\rm 1e(h)}}\, b_{\rm N} b_{\rm e(h)} \langle S \rangle^2
\left[
\frac{B_{\rm V} - \omega/\gamma_{\rm N}}{b_{\rm e(h)}^2 \langle S \rangle^2 + \left(B_{\rm V} - \omega/\gamma_{\rm N}\right)^2}
+
\frac{B_{\rm V} + \omega/\gamma_{\rm N}}{b_{\rm e(h)}^2 \langle S \rangle^2 + \left(B_{\rm V} + \omega/\gamma_{\rm N}\right)^2}
\right],
\label{eq:nuclearfield}
\end{equation}
\end{widetext}
where
\begin{equation}
\langle S \rangle
=
\frac{\langle S_0 \rangle}
{\sqrt{1 + \dfrac{(B + B_{\rm N})^2}{\Delta_{\rm H}^2}}}
\label{eq:s}
\end{equation}
is the amplitude of the oscillating average spin of a charge carrier under Hanle effect conditions. $b_{\rm N}$ and $b_{\rm e(h)}$ are the Overhauser and Knight field values for fully polarized nuclear and electron (hole) spins, $T_1$ and $T_{\rm 1e(h)}$ denote the time constants characterizing the nuclear spin-lattice relaxation and the dynamic polarization of nuclear spins by electrons (holes), $\omega$ is the oscillation frequency of the Knight field, corresponding to the modulation frequency of the EOM ($\omega=2\pi f_{\rm mod}$), and $\gamma_{\rm N}$ is the nuclear gyromagnetic ratio. The numerical solution of the nonlinear equation~\eqref{eq:nuclearfield} provides the nuclear field $B_{\rm N}$ as  function of the external field $B_{\rm V}$. This allows us to model the shape of the Hanle curve using Eq.~\eqref{eq:Hanle_formula}, where the external magnetic field $B_{\rm V}$ is replaced by the total field $B_{\rm V} + B_{\rm N}$.

Figures~\ref{fig:Hanle_Cooling_Theory}(a,d) show the Hanle curves, plotted as blue lines, calculated for charge carriers with a negative (hole) $g$-factor and a positive (electron) $g$-factor for $f_{\rm mod}=200$~kHz. Note that the sign of the Overhauser field depends on the sign of the charge carrier $g$-factor and on whether the external magnetic field is smaller or larger the NMR resonance condition. We illustrate that by Figs.~\ref{fig:Hanle_Cooling_Theory}(b,e), where $B_{\rm N}$ is plotted as function of $B_{\rm V}$. One can see that in a nuclear spin system undergoing resonant cooling via an oscillating Knight field, the Overhauser field's polarity is determined by the frequency detuning, resulting in a sign change at the resonance point~\cite{meier1984optical}.  


In the case when the external field, $B_{\rm V}$, is smaller than the NMR resonance field, $B_{\rm res} = 2\pi f_{\rm mod}/\gamma_{\rm N}$, i.e.  $B_{\rm V}<B_{\rm res}$, the Overhauser field for $^{207}$Pb has a negative sign, being antiparallel to $B_{\rm V}$ for holes with $g_{\rm h}<0$. As a result, the total effective field acting on the hole $B_{\rm V} + B_{\rm N}$ is reduced, see Fig.~\ref{fig:Hanle_Cooling_Theory}(c), where the dependence of $B_{\rm V} + B_{\rm N}$ on $B_{\rm V}$ is plotted. Therefore, the depolarizing effect becomes smaller. 
Passing through the resonance ($B_{\rm V}=B_{\rm res}$), where the Overhauser field is zero, a sign reversal of $B_{\rm N}$ occurs, see Fig.~\ref{fig:Hanle_Cooling_Theory}(b). For the external field exceeding the resonance field, $B_{\rm V}>B_{\rm res}$, the Overhauser field for holes adds to the external field, resulting in a stronger decrease in the degree of optical orientation in the Hanle curve. This gives rise to a step-like resonance feature in the Hanle curve, see Fig.~\ref{fig:Hanle_Cooling_Theory}(a).

For electrons with a positive $g$-factor, the appearance is just opposite to that of the holes, compare Figs.~\ref{fig:Hanle_Cooling_Theory}(e,f) with Figs.~\ref{fig:Hanle_Cooling_Theory}(b,c). This results in the derivative-like resonance feature in the Hanle curve, Fig.~\ref{fig:Hanle_Cooling_Theory}(d). Therefore, the shape of the resonance feature gives direct information about the carrier $g$-factor sign, which allows us to identify for MAPbI$_3$ crystals the carrier type contributing to the resonant nuclear cooling. 

Taking this into account, we conclude that the resonant cooling observed in the Hanle curves in Fig.~\ref{fig:Hanle_Cooling} is associated with carriers having a negative $g$ factor, i.e. with holes interacting with the $^{207}$Pb nuclei. Since the half-width of the Hanle curve of the weakly localized holes (0.4 mT) is much smaller than the resonance fields corresponding to the used modulation frequencies, we attribute the observed resonant cooling of nuclear spins to strongly localized holes with a half-width of 37 mT. This conclusion is further supported by the theoretical Hanle curve presented in Fig.~\ref{fig:Hanle_Cooling_Theory}(a). Accordingly, as fit parameter we used $\Delta_{\rm H}=37$~mT, along with $b_{\rm N}T_1/T_{\rm 1e(h)}=200$~mT and $b_{\rm e(h)}=20$~mT. For direct comparison with the experimental data, the calculated Hanle curve (blue line) was supplemented by the contributions from electrons ($\Delta_{\rm H}=9$~mT) and weakly localized holes ($\Delta_{\rm H}=0.4$~mT), described by Eq.~(2) (red dashed line in Fig.~\ref{fig:Hanle_Cooling_Theory}(a)). The Knight field substantially exceeds the local field of the nuclear dipole–dipole interactions, $B_{\rm L} = 0.25$ mT~\cite{kotur2026dynamic}, indicating that the resonant cooling of the lead spins occurs in the strong Knight-field regime, which has been considered theoretically in Ref.~\onlinecite{kavokin2026resonant}. The broad, high-amplitude resonances observed experimentally are consistent with the predictions of this model.

\section{Conclusions}


Resonant cooling of nuclear spins by optically-oriented holes has been demonstrated for lead halide perovskite crystals using MAPbI$_3$ crystals as representative material. The Hanle curves measured at $E_{\rm det}=1.628$~eV reflect the resonant optical cooling of the nuclear spins, as evidenced by a shift of the resonance field toward higher magnetic fields with increasing modulation frequency. No change in the Hanle curve shape is observed upon in-plane rotation of the sample from $0^\circ$ to $45^\circ$, which is consistent with an isotropic hyperfine interaction. This behavior is expected for the $^{207}$Pb nuclei with spin $I=1/2$, which are not affected by quadrupolar interactions due to the absence of a nuclear quadrupole moment. 
We conclude from the shape of the resonance feature in the Hanle curve, that $^{207}$Pb nuclei are cooled by carriers with a negative $g$-factor, consistent with hole-mediated processes. These results demonstrate that the rich spectrum of spin-dependent phenomena induced by the hyperfine interaction of carriers with nuclear spins, as established for common semiconductors, can be expected as well in the lead halide perovskite semiconductor material class.

\section{Experimental Section}
\label{sec:experiment}

\subsection*{Sample}
We study a methylammonium lead iodide (MAPbI$_3$) perovskite single crystal with a thickness of approximately 30 $\mu$m (sample code M2-7), synthesized using the inverse temperature crystallization method~\citep{chen2019single, alsalloum2020low}. At room temperature, the MAPbI$_3$ single crystal shows a tetragonal crystal structure, with the $[001]$ direction normal to the sample surface. Under cryogenic conditions, the crystal undergoes a phase transition to a orthorhombic structure. The electrons and holes in MAPbI$_3$ have opposite $g$-factor signs, with $g_{\rm e}=+2.83$ and $g_{\rm h}=-0.54$~\cite{kopteva2025optical}. Further information on the optical and spin properties of MAPbI$_3$ crystals can be found in Refs~\cite{kotur2026mapi, kopteva2025optical}.

\subsection*{Experimental setup}
The sample was mounted in the variable temperature insert of a helium bath cryostat, allowing temperatures from 1.6 K up to room temperature. The base temperature of $T=1.6$~K used in this study corresponds to conditions in which the sample is immersed in superfluid helium. Optical excitation was provided by a continuous-wave Ti:Sapphire laser with the photon energy of $E_{\rm exc} = 1.722$~eV, tuned to a value above the MAPbI$_3$ band gap energy of $1.652$~eV~\citep{kopteva2025optical}. The laser beam was circularly polarized, with helicity alternated between $\sigma^+$ and $\sigma^-$ using an electro-optical modulator (EOM). It was directed along the $[001]$ crystallographic axis, $\mathbf{k}\parallel[001]$. A pulse generator provided the trigger signal for the EOM, enabling variation of the modulation frequency $f_{\rm mod}$ in the range from 10 to 400~kHz. 

\subsection*{Photoluminescence (PL) Measurement}
The photoluminescence (PL) of the crystal was measured with an avalanche photodiode (APD) connected to a 0.5-meter spectrometer. A polarization analyzer in the detection channel, consisting of a $\lambda/4$ wave plate and a Glan–Taylor prism, was used to analyze the optical orientation degree of the PL and to simultaneously detect the $I^{++}$ and $I^{-+}$ intensities of the $\sigma^+$ polarized PL component, using a two-channel photon counter, synchronized with the EOM in the excitation path. The optical orientation degree is defined as
\begin{equation}
P_{\rm oo}=\frac{I^{++}-I^{-+}}{I^{++}+I^{-+}}.
\label{eq:optical_orientation_degree}
\end{equation}
Here $I$ is the photoluminescence intensity, for which the first upper index indicates the circular polarization of excitation and the second upper index the polarization of emission. An external electromagnet provided a magnetic field up to $B_{\rm V}=125$~mT applied in the Voigt geometry, $\mathbf{B}_{\rm V} \perp \mathbf{k}$ ($x$ axis).

\section*{Author information}
\subsection*{Corresponding Authors}
Mladen Kotur - \textit{Experimentelle Physik 2, Technische
Universität Dortmund, 44227 Dortmund, Germany};\\
\href{https://orcid.org/0000-0003-0865-0393}{orcid.org/0000-0002-2569-5051};\\
Email: \href{mailto:mladen.kotur@tu-dortmund.de}{mladen.kotur@tu-dortmund.de}

Dmitri R. Yakovlev - \textit{Experimentelle Physik 2, Technische
Universität Dortmund, 44227 Dortmund, Germany};\\
\href{https://orcid.org/0000-0001-7349-2745}{orcid.org/0000-0001-7349-2745};\\
Email: \href{mailto:dmitri.yakovlev@tu-dortmund.de}{dmitri.yakovlev@tu-dortmund.de}

\subsection*{Authors}
Nataliia E. Kopteva - \textit{Experimentelle Physik 2, Technische Universität Dortmund, 44227 Dortmund, Germany};\\
\href{https://orcid.org/0000-0003-0865-0393}{orcid.org/0000-0003-0865-0393}

Bekir Turedi - \textit{Laboratory of Inorganic Chemistry, Department of Chemistry and Applied Biosciences,  ETH Z\"{u}rich, CH-8093 Z\"{u}rich, Switzerland}; \textit{Laboratory for Thin Films and Photovoltaics, Empa-Swiss Federal Laboratories for Materials Science and Technology, CH-8600 D\"{u}bendorf, Switzerland};\\
\href{https://orcid.org/0000-0003-2208-0737}{orcid.org/0000-0003-2208-0737}

Maksym V. Kovalenko - \textit{Laboratory of Inorganic Chemistry, Department of Chemistry and Applied Biosciences,  ETH Z\"{u}rich, CH-8093 Z\"{u}rich, Switzerland}; \textit{Laboratory for Thin Films and Photovoltaics, Empa-Swiss Federal Laboratories for Materials Science and Technology, CH-8600 D\"{u}bendorf, Switzerland};\\
\href{https://orcid.org/0000-0002-6396-8938}{orcid.org/0000-0002-6396-8938}

Manfred Bayer - \textit{Experimentelle Physik 2, Technische
Universität Dortmund, 44227 Dortmund, Germany}; \textit{Research Center FEMS, Technische Universit\"at Dortmund, 44227 Dortmund, Germany};\\
\href{https://orcid.org/0000-0002-0893-5949}{orcid.org/0000-0002-0893-5949}

\subsection*{Notes}

The authors declare no conflict of interest.

\section*{Acknowledgments} 
The authors thank K.V. Kavokin for fruitful discussions. M.K. and M.B. acknowledge the support of the BMBF project QR.N (Contract No.16KIS2201). D.R.Y. acknowledges support from the Deutsche Forschungsgemeinschaft via the SPP2196 Priority Program (Project YA 65/28-1, No. 527080192). N.E.K. acknowledges support from the Deutsche Forschungsgemeinschaft (Project KO 7298/1-1, No. 552699366). The work at ETH Z\"urich (B.T. and M.V.K.) was financially supported by the Swiss National Science Foundation (grant agreement 200020E 217589, funded through the DFG-SNSF bilateral program) and by ETH Z\"urich through ETH+ Project SynMatLab.

%


\end{document}